\documentclass[12pt]{report} 

\usepackage[a4paper,portrait,total={17cm,25cm}]{geometry}

\usepackage{authblk}
\title{From $k+1$ to $k$ heads\\the descriptive trade-off is non-recursive\\
{\small\textit{Proceedings of DCFS 2004}, pp.213–224.}} 
\author{Christos Kapoutsis}
\affil{Computer Science and Artificial Intelligence Laboratory\\
    Massachusetts Institute of Technology\\
    \texttt{cak@mit.edu}}
\date{}

\usepackage{amsmath}
\usepackage{amsthm}
\usepackage{rotating}

\theoremstyle{plain}
\newtheorem{theorem}{Theorem}[section]
\newtheorem{lemma}[theorem]{Lemma}

\theoremstyle{definition}
\newtheorem{definition}[theorem]{Definition}

\newtheorem{xample}[theorem]{Example}

\newcommand{\modelofcomputation}[1]{\text{\scshape\mdseries\rmfamily #1}}

\newcommand{\odfa}{\modelofcomputation{{{\footnotesize 1}dfa}}}

\newcommand{\tdfa}{\modelofcomputation{{{\footnotesize 2}dfa}}}
\newcommand{\tnfa}{\modelofcomputation{{{\footnotesize 2}nfa}}}

\newcommand{\utdfa}{\tdfa^{\textup{u}}}
\newcommand{\utnfa}{\tnfa^{\textup{u}}}

\newcommand{\mtdfa}{\tdfa_*}
\newcommand{\mtnfa}{\tnfa_*}
\newcommand{\mutdfa}{\tdfa^{\textup{u}}_*}

\newcommand{\dca}{\modelofcomputation{{dca}}}
\newcommand{\mdca}{\dca_*}

\newcommand{\dpa}{\modelofcomputation{{\footnotesize 1}dpa}}

\newcommand{\tm}{\modelofcomputation{tm}}

\newcommand{\complexityclass}[1]{\text{\upshape\bfseries\rmfamily #1}}

\renewcommand{\l}{\complexityclass{L}} 

\newcommand{\complexityproblem}[1]{\text{\upshape\mdseries\sffamily #1}}

\newcommand{\ess}{\complexityproblem{H}}
\newcommand{\looping}{\overline{\complexityproblem{HALTING}}}

\newcommand{\ggG}{\varGamma}
\newcommand{\ggD}{\varDelta}
\newcommand{\ggP}{\varPi}
\newcommand{\ggS}{\varSigma}

\newcommand{\set}[1]{\{#1\}}

\newcommand{\amp}{\;\;\&\;\;}
\newcommand{\mach}[1]{{(#1)}}
\newcommand{\lang}[1]{{(#1)}}
\newcommand{\desc}[1]{{\langle#1\rangle}}

\newcommand{\s}{^*}

\newcommand{\nats}{{\mathbb{N}}}

\newcommand{\lend}{\vdash}
\newcommand{\rend}{\dashv}

\newcommand{\size}{\textup{size}}
\newcommand{\eff}{f}

\newcommand{\PP}{\ggP}
\newcommand{\PPy}{\ggP_{\textup{yes}}}
\newcommand{\PPn}{\ggP_{\textup{no}}}
\newcommand{\eot}{\complexityproblem{E}}
\newcommand{\eoty}{\eot_{\textup{yes}}}
\newcommand{\eotn}{\eot_{\textup{no}}}

\newcommand{\bla}{\sqcup}
\newcommand{\dotz}{\dot{0}}
\newcommand{\doto}{\dot{1}}

\usepackage{amssymb}

\begin{document}
\maketitle

\begin{abstract}
We prove that no recursive function can upper bound the increase in
the size of description when a two-way deterministic finite automaton
with $k+1$ heads is replaced by an equivalent two-way deterministic 
finite automaton with $k$ heads. This is true for all $k$, and remains
true if the  automata are unary and/or nondeterministic.
\end{abstract}

\section{Introduction}
\label{sec:introduction}
Let $\tdfa_k$ stand for `two-way deterministic finite automaton with
$k$ heads' and $\lang{\tdfa_k}$ for the set of languages recognizable
by such automata. It is well known that the sequence
$$
\lang{\tdfa_1},\; 
\lang{\tdfa_2},\; 
\dots, \;
\lang{\tdfa_k},\;
\dots
$$
starts with the regular languages~\cite{rasc59,sh59}, is strictly
increasing~\cite{mo76,mo77}, and eventually covers the entire class
$\l$ of languages decidable by Turing machines ($\tm$s) in logarithmic
space~\cite{ha72}. Hence, for any $L\in\l$ and for $k\s$ the smallest
$k$ for which $L\in\lang{\tdfa_k}$, each one of the following types of
machines
$$
\tdfa_{k\s},\;
\tdfa_{k\s+1},\; 
\tdfa_{k\s+2},\;
\ldots\quad,\;
\text{logarithmic-space }\tm,
$$
suggests a different string to `describe' $L$: a smallest (description
of a) machine of this type that recognizes $L$. It is then conceivable
that stronger types of machines may provide more succinct descriptions
than weaker types, making it natural to ask how much more verbose the
weaker types can be, in general.

Recently, Kutrib~\cite{ku03} proved that, $\tdfa_k$s can be both
non-recursively more verbose than logarithmic-space $\tm$s and
non-recursively more succinct than $\tdfa_1$s. That is, he showed that
no recursive function can upper bound the increase in the size of
description when we replace a minimal (description of a)
logarithmic-space Turing machine with (a description of) an equivalent
$\tdfa_k$, for any $k\geq 1$. Similarly, on the other end of the
hierarchy, the increase in the size of description when a minimal
(description of a) $\tdfa_k$ is replaced by (a description of) an
equivalent $\tdfa_1$ admits no recursive upper bound, either, for any
$k\geq 2$.

Kutrib went on to conjecture that the trade-off is non-recursive even
when we change from a minimal $\tdfa_{k+1}$ into an equivalent
$\tdfa_k$, for all $k$. In addition, he posed the question whether the
same is true when we restrict our attention to unary automata of this
type ---for this case he found huge recursive lower bounds but
refrained from guessing that no recursive upper bound exists, because
of the real-time one-way cellular automata precedent (for these
automata, non-recursive trade-offs for arbitrary languages reduce to
recursive trade-offs for unary languages~\cite{ma02}).

We resolve these questions by showing that non-recursive blow-ups are
possible both in the general and in the unary case. Moreover, if we
switch our attention to nondeterministic automata, the blow-ups can
still be non-recursive, again both in the general and in the unary
case. Section~\ref{sec:framework} describes the framework of this
study in more detail and gives an outline of the proof, whose two
pieces are presented in Sections~\ref{sec:from-looping-to-eot}
and~\ref{sec:from-eot-to-ess_k}. The discussion focuses on the general
deterministic case, which is the most natural one; when appropriate, a
final note in each section addresses the unary and/or nondeterministic
cases.

\paragraph{Related work.}
The study of the relative succinctness of descriptional systems of
different power seems to have started with Stearns~\cite{st67}, who
gave a recursive upper bound for the size of the smallest
deterministic finite automaton ($\odfa$) that recognizes the language
of a given deterministic pushdown automaton ($\dpa$) whenever this
language is indeed regular. Following this, Meyer and
Fischer~\cite{mefi71} showed no such bound exists if we compare
$\odfa$s to context-free grammars ---the first proof of a
non-recursive trade-off. Later, Valiant~\cite{va76} proved the same
for the comparison between $\dpa$s and unambiguous context-free
grammars, while Schmidt and Szymanski~\cite{scsz77} repeated this for
the comparison between the latter and (general) context-free grammars.
Finally, Hartmanis~\cite{ha79} simplified many of the proofs by noting
the relation of the recursiveness of a trade-off to the
recognizability of the corresponding inadequacy problem
(cf.~Lemma~\ref{lem:basic}); he also proved non-recursive trade-offs
even for comparisons where the nondeterminstic descriptors from above
are accompanied by proofs that their languages are deterministic or
unambiguous. It is this discussion that Kutrib~\cite{ku03} has
transferred to the domain of multihead finite automata.

For a detailed overview of the subject, along with a comprehensive
survey of the broader field of descriptive complexity, the reader is
referred to~\cite{goea02}.

\section{The framework}
\label{sec:framework}

We denote the set of positive integers by $\nats$. For $n,a\in\nats$,
we write $\lg_a n$ to denote $\lfloor\log_a n\rfloor$. The set of all
finite strings over an alphabet $\ggG$ is denoted by $\ggG\s$.  

A (\emph{promise}) \emph{problem} over $\ggG$ is any pair
$\PP=(\PPy,\PPn)$ where $\PPy$, $\PPn$ are disjoint subsets of
$\ggG\s$. A $\tm$ \emph{recognizes} $\PP$ if it accepts all $x\in\PPy$
and rejects all $x\in\PPn$ ---possibly by looping. If some $\tm$
recognizes $\PP$, we say $\PP$ is \emph{recursively enumerable} or
(following~\cite{si96}) (\emph{Turing-}) \emph{recognizable}.  For
$\PP'$ also a problem over $\ggG$, we write $\ggP\leq\ggP'$ and say
$\PP$ \emph{reduces to} $\PP'$ iff there is a $\tm$ that, on input
$x\in\ggG\s$, eventually halts with an output $y$ such that
$
x\in\PPy \implies y\in\PPy'
$
and
$
x\in\PPn \implies y\in\PPn'
$.
If some unrecognizable $\PP$ reduces to $\PP'$, then $\PP'$ is also
unrecognizable~\cite{houl79,si96}.

If $\PPn=\overline{\PPy}$, then $\PP$ is also called a \emph{language}
and is adequately described by $\PPy$ alone. If in addition $\PPy$
contains exactly all sufficiently long strings for some interpretation
$l$ of `sufficiently long',
$
\PPy = \set{ x\in\ggG\s \mid \text{length($x$)} \geq l }
$,
then we say $\PP$ is a \emph{threshold language}.

\subsection{Multihead automata}
\label{sec:multihead-automata}

A \emph{two-way deterministic finite automaton with $k$ heads} (a
$\tdfa_k$) \emph{over alphabet $\ggS$} consists of a finite state
control, $k$ read-only heads, and a tape that can represent the
symbols of $\ggS\neq\emptyset$ and the extra endmarking symbols
$\lend$ and~$\rend$. An input $x\in\ggS$ is presented on the tape
surrounded by the endmarkers.  The automaton starts at a designated
\emph{start state}, all heads reading the left endmarker $\lend$. At
every step, the $k$ symbols under the heads are read; based on this
information and the current state, the automaton decides what its next
state should be, which one head it should move, and whether to move it
left or right; it then simultaneously changes its state and moves the
chosen head accordingly. No head can move past an endmarker and the
automaton cannot sense when two of its heads are on the same tape
cell. The input is \emph{accepted} if the machine ever reaches a
designated \emph{final state} and halts. The \emph{language} of the
automaton is exactly the set of strings over $\ggS$ that it
accepts. We say the automaton is \emph{terminating} if it halts on all
inputs.

By a \emph{multihead two-way deterministic finite automaton} (a
$\mtdfa$), we mean a $\tdfa_k$ with any number of heads~$k$. A
\emph{description} of such an automaton is any string that encodes it,
under some arbitrary (but fixed) reasonable encoding of $\mtdfa$s into
finite strings over some alphabet~$\ggD$. The \emph{size} of such a
description $z\in\ggD\s$ is the value $\size(z)$, where
\,\mbox{$\size\!\!:\!\ggD\s\to\nats$}\,\, is some arbitrary (but
fixed) computable function with the property that some $\tm$ can, on
input $k,s\in\nats$, produce the list of all descriptions of
$\tdfa_k$s that have size at most~$s$.

We write $\mach{z}$ for the $\mtdfa$ encoded by a description $z$, and
$\lang{M}$ for the language of a $\mtdfa$ $M$; the set of all
descriptions of $M$ (which may not be a singleton) is denoted by
$\desc{M}$. The collection of the languages of all $\tdfa_k$s [all
$\mtdfa$s] is denoted by $\lang{\tdfa_k}$ [by $\lang{\mtdfa}$]. The
collection of the descriptions of all $\tdfa_k$s [all $\mtdfa$s] is
denoted by $\desc{\tdfa_k}$ [by $\desc{\mtdfa}$]. Two $\mtdfa$s are
\emph{equivalent} if their languages are identical; two descriptions
$z,z'\in\desc{\mtdfa}$ are equivalent if $(z)$, $(z')$ are.

\paragraph{The unary and nondeterministic cases.}
If $\ggS$ in the definition of a $\tdfa_k$ contains only one symbol,
we say the automaton is \emph{unary} (a $\utdfa_k$). If at every step
more than one next moves are possible, we say the automaton is
\emph{nondeterministic} (a $\tnfa_k$; or a $\utnfa_k$, if the alphabet
is unary). The sets of languages $\lang{\utdfa_k}$, $\lang{\mtnfa}$,
etc.\ and descriptions $\desc{\utdfa_k}$, $\desc{\mtnfa}$, etc.\ are
defined analogously.

\subsection{Decreasing the number of heads}
\label{sec:decr-numb-heads}

It is known~\cite{mo76,mo77} that $\tdfa_{k+1}$s are more powerful
than $\tdfa_k$s, in the sense that 
\begin{equation} \label{eq:k+1-better-than-k}
\lang{\tdfa_{k}} 
\subsetneqq 
\lang{\tdfa_{k+1}},
\end{equation}
for all $k$. This suggests two natural discussions.

\textsc{i.}~Since $\lang{\tdfa_k}\subseteq\lang{\tdfa_{k+1}}$, an
$L\in\lang{\tdfa_k}$ can be `described' not only by descriptions of
$\tdfa_k$s that recognize it, but also by descriptions of
$\tdfa_{k+1}$s. Conceivably, the minimal descriptions of the former
type are larger than the minimal descriptions of the latter type, in
which case to change from a minimal $\tdfa_{k+1}$ description of $L$
to a $\tdfa_k$ description of $L$ is to trade an increase in the size
of the description for a decrease in the number of heads. It is then
natural to ask for an upper bound for this increase.
\begin{definition} \label{def:upper-bound}
If \mbox{$\eff_k:\nats\to\nats$} is such that 
\begin{quote}
for all $s\in\nats$ and all $z\in\desc{\tdfa_{k+1}}$ of size at most
$s$: if $z$ has an equivalent $z'\in\desc{\tdfa_k}$, then it has one
of size at most~$\eff_k(s)$,
\end{quote}
then we say that $\eff_k$ is an upper bound for the increase in the
size of description when a $\tdfa_{k+1}$ is replaced by an
equivalent~$\tdfa_k$.
\end{definition}
\noindent
So, if $\eff_k$ is such an upper bound and $M$ is a $\tdfa_{k+1}$ that
has equivalent $\tdfa_k$s, it is possible to find one with a
description no larger than $\eff_k(s)$, where $s$ is the size of any
description of $M$. To define such an $\eff_k$ is, of course,
very easy; what is questionable is whether we can define one that is
also recursive. If this is not possible, we say that \emph{the
$($descriptive$)$ trade-off from $k+1$ to $k$ heads is non-recursive
$($for $\mtdfa$s$)$}, or that $\tdfa_{k+1}$s \emph{can be
non-recursively more succinct than} $\tdfa_k$s.

\textsc{ii.}~Since
$\lang{\tdfa_k}\neq\lang{\tdfa_{k+1}}$, a $\tdfa_{k+1}$ may or may not
have an equivalent $\tdfa_k$, so that it is natural to consider the
associated computational problem: ``Given a description of a
$\tdfa_{k+1}$ $M$, is it true that $M$ has no equivalent $\tdfa_k$?''
\begin{definition}
$ 
\ess_k = \bigl\{ 
   z\in\desc{\tdfa_{k+1}} 
   \mid 
   \neg\bigl( \exists z'\in\desc{\tdfa_k} \bigr)
   \bigl(\textup{$z$, $z'$ are equivalent} \bigr)
\bigr\}.
$
\end{definition}
\noindent
In a more intuitive dialect, $\ess_k$ is the problem: ``Given a
$\tdfa_{k+1}$~$M$, is it true that $M$ can't perform the same task
with only $k$ of its heads?'' We are interested in the question
whether this problem is recognizable, in which case we say that
\emph{the inadequacy of $k$ heads is recognizable $($for
$\tdfa_{k+1}$s$)$}.

As first noted in~\cite{ha79}, discussions (\textsc{i}) and
(\textsc{ii}) are not unrelated (see also~\cite{ku03}).

\begin{lemma}\label{lem:basic}
If the trade-off from $k+1$ to $k$ heads is recursive, then the
inadequacy of $k$ heads is recognizable.
\end{lemma}

\paragraph{The unary and nondeterministic cases.}
That $k+1$ heads are better than $k$ has also been established for the
case where the automata are unary and/or
nondeterministic~\cite{mo77,mo80}. That is, similarly
to~(\ref{eq:k+1-better-than-k}), we also have
$$
\lang{\utdfa_k}\subsetneqq\lang{\utdfa_{k+1}}, \qquad
\lang{\tnfa_k} \subsetneqq\lang{\tnfa_{k+1}},  \qquad
\lang{\utnfa_k}\subsetneqq\lang{\utnfa_{k+1}},
$$
for all $k$.  As a result, for each one of the three types of machines
we can repeat the preceding discussions, asking again
(\textsc{i})~whether the trade-off from $k+1$ to $k$ heads is
recursive, and (\textsc{ii})~whether the inadequacy of $k$ heads is
recognizable. Lemma~\ref{lem:basic} can then be shown anew (with the
same proof).

\subsection{Proof outline}
\label{sec:outline-proof}

Fix any $k\geq 1$. Our goal is to prove that $\tdfa_{k+1}$s can be
non-recursively more succinct than $\tdfa_k$s. By
Lemma~\ref{lem:basic}, it is enough to show that $\ess_k$ is
unrecognizable. We do this by a reduction from the complement of the
halting problem
$$
\looping \leq \ess_k,
$$
where 
$
\looping = \{ 
   z\in\{0,1\}\s 
   \mid 
   \text{$z$ encodes a $\tm$ that loops on $z$}
\}
$
is known to be unrecognizable~\cite{houl79,si96}. What we actually
show is two separate reductions:
\begin{equation} \label{eq:two-reductions}
\looping \leq \eot \leq \ess_k,
\end{equation}
where $\eot$ is the promise problem: ``Given a description of a
terminating $\utdfa_2$ whose language is either empty or threshold,
check that it is empty.'' In formal dialect, $\eot=(\eoty,\eotn)$,
where
\begin{align*}
\eoty
&= 
\{ z\in\desc{\utdfa_2} 
   \mid 
   \textup{$\mach{z}$ is terminating 
   and
   $\lang{\mach{z}}=\emptyset$} 
\},
\\
\eotn
&= 
\{ z\in\desc{\utdfa_2} 
   \mid 
   \textup{$\mach{z}$ is terminating 
   and 
   $\lang{\mach{z}}$ is a threshold language}
\}.
\end{align*} 
We present the two reductions of~(\ref{eq:two-reductions}) separately,
in Sections~\ref{sec:from-looping-to-eot}
and~\ref{sec:from-eot-to-ess_k}. The first reduction needs a
special type of automaton, which we introduce in the next section.

\paragraph{The unary and nondeterministic cases.}
Lemma~\ref{lem:basic} again suggests that we need only prove the
inadequacy of $k$ heads is unrecognizable, for each one of the three
types of automata. We do this with the same two reductions
of~(\ref{eq:two-reductions}). Only the second one needs to be modified.

\subsection{Multicounter automata}
\label{sec:bound-mult-autom}

A \emph{deterministic automaton with $k$ $($bounded$)$
counters}\footnote{In this definition the reader will
probably almost recognize the \emph{bounded counting automata}
of~\cite{risp72}, the \emph{two-way counter automata} and the
\emph{register machines} of~\cite{mo80}, or other automata from
elsewhere. The present model is nicely simpler, in the sense that the
input is directly taken to be the upper bound for the counters, saving
us the redundant (in this case) notion of an input tape or an input
register.}  ($\dca_k$) consists of a finite state control and $k$
counters, each of which can store a nonnegative integer. The input to
the automaton is a nonegative upper bound $n$ for its counters. The
automaton starts at a designated \emph{start state} with all its
counters set to~0. At every step, based on its current state, the
automaton decides which counter it should act upon and whether it
should increase it or decrease it. Then the action is
\emph{attempted}. An attempt to decrease fails iff the counter already
contains~0; an attempt to increase fails iff the counter already
contains~$n$. A failed attempt leaves the counter contents intact; a
successful attempt updates the counter contents accordingly. Based on
its current state and on whether the attempt succeeded or not, the
automaton selects a new state and moves to it. The input is
\emph{accepted} if the machine ever enters a designated \emph{final
state}. The \emph{language} of the automaton is exactly the set of
nonnegative integers that it accepts.  By a \emph{deterministic
multicounter automaton} (a $\mdca$) we mean a $\dca_k$ with any number
of counters~$k$.

The new automata are just another, more convenient way to look
at~$\mutdfa$s.

\begin{lemma} \label{lem:dca-vs-utdfa}
Every $\utdfa_k$~$M$ has a $\dca_k$~$M'$ such that 
$
\lang{M'} = \{ n+1 \mid 0^n \in \lang{M} \}
$. 
Conversely, 
every $\dca_k$ $M$ has a $\utdfa_k$ $M'$ with
$
\lang{M'} = \{ 0^{n-1} \mid n\in\lang{M} \amp n\neq 0 \}
$.
\end{lemma}

The computing capabilities of $\mdca$s with 2 or more counters are
quite significant, as explained in the Appendix. We will be using
these capabilities in the next section.

\section{The first reduction}
\label{sec:from-looping-to-eot}

In this section, our goal is to find an algorithm that, on input a
decription $z$ of a $\tm$ $M$ produces a description $z'$ of a
terminating $\utdfa_2$ $M'$ such that
\begin{equation} \label{eq:reduction-condition-1}
\begin{split}
M\text{ loops on input }z 
&\implies
\lang{M'}=\emptyset,
\\
M\text{ halts on input }z 
&\implies
\lang{M'}\text{ is a threshold language}.
\end{split}
\end{equation}
We will be calling a machine ($\tm$, $\mdca$, or $\mutdfa$)
\emph{good} if it is terminating and its language
satisfies~(\ref{eq:reduction-condition-1}) when it replaces
$\lang{M'}$. For example, $(z')$ should be good.
 
On its way to $z'$, our algorithm will construct descriptions of three
other machines, in this order: a description $z_A$ of a $\tm$ $A$, a
description $z_B$ of a $\dca_3$ $B$, and  a description $z_C$ of a
$\dca_2$ $C$.  In the sequence $M$, $A$, $B$, $C$, $M'$, each machine
after $M$ will be defined in terms of the previous one and will be
good. The reader will probably recognize in our constructions the
ideas of~\cite{wa57} and~\cite{mi61} (also found
in~\cite{mi67,houl79}).

\paragraph{The first machine.}
$A$ is a $\tm$ with one tape, infinite in both directions; the tape
alphabet is $\{\bla,0,1,\dotz,\doto\}$, while the input alphabet is
$\{0\}$. On input $0^n$, $A$ starts with tape contents
$$
\cdots\bla\bla\bla 
\underbrace{000\cdots00}_{n\text{ times}} 
\bla\bla\bla\cdots
$$
and its head reading the leftmost~0 (or a $\bla$, if $n=0$). It then
computes:

\smallskip
\quad 
1.~For all $x\in\{0,1\}^n$, from $0^n$ up to $1^n$: \nopagebreak

\quad\qquad 
--- if $x$ encodes a halting computation 
history of~$M$ on~$z$, accept. \nopagebreak

\quad
2.~Reject.

\smallskip\noindent   
The check inside the loop presupposes some fixed
reasonable encoding of sequences of configurations of $M$ into binary
strings, with the additional property: \emph{if $w$ encodes a
computation history, then every string of the form $w0^*$ encodes the
same history}. Note that $A$ can easily perform this check using the
extra dotted symbols, never writing a non-blank symbol on a~$\bla$, or
a $\bla$ on a non-blank symbol.

\paragraph{The second machine.}
$B$ is a $\dca_3$ that, on input $n\geq 30$, simulates the behaviour
of $A$ on input $0^{\lg_5 \lg_{30} n}$; on input $n<30$, $B$ just
rejects.\footnote{The strange length $\lg_5\lg_{30}n$ of the input of
$A$'s simulated computation is chosen as a function of $n$ that is
(i)~\emph{computable} by a $\dca_3$ and (ii)~\emph{increasing}, but
also (iii)~\emph{small enough}. Goodness of $B$ bases on~(ii), while
(iii)~facilitates the simulation performed by $C$ in the next section.}

To explain $B$'s behavior, let $L$, $R$, $J$ be its three
counters. $J$ is auxiliary, used for performing operations on $L$ and
$R$, while $L$ and $R$ together encode tape configurations of $A$. To
see the encoding, consider the following example of a configuration:
\begin{center}
\footnotesize
$
\begin{array}{cccccccccccccc}
\cdots& l_4& l_3& l_2  & l_1  & l_0  &  r_0  & r_1  & r_2  & r_3  & r_4 & r_5& r_6&\cdots \\ \hline
\cdots&\bla&\bla&\times&\times&\times&\times&\times&\times&\times&\times&\bla&\bla&\cdots \\ \hline
      &    &    &      &      &      &\uparrow&    &      &      &    &    &  
\end{array}
$
\end{center}
(here {\footnotesize$\times$} stands for any non-blank symbol,
{\footnotesize$\uparrow$} shows the head position). Mapping symbols
$\bla$, $\doto$, $\dotz$, $1$ and $0$ to numbers $0$, $1$, $2$, $3$
and $4$, respectively, we get each tape cell map to a digit of the
5-ary numbering system. Then, the head position splits the tape into
two portions, which define the integers
$$
l = \sum_{i=0}^\infty l_i \cdot 5^i
\qquad\text{and}\qquad
r = \sum_{r=0}^\infty r_i \cdot 5^i
$$
(note that the cell under the head contributes to $r$; also, the sums
are finite exactly because $\bla$ maps to~0). These two values are
kept in $L$ and $R$, respectively.

More specifically, on input $n$, $B$ starts with a two-part
initialization phase. \emph{First}, it computes $\lg_{30} n$ into $J$,
leaving $0$'s in $L$ and $R$ (if $n=0$, $B$ rejects). \emph{Second},
it computes into $R$ the value $5^m-1$, where $m=\lg_5\lg_{30}n$,
leaving $0$ in $L$ (this is only if $J\geq 1$, that is if $n\geq 30$;
otherwise, $n<30$ and $B$ rejects). This completes the initialization,
with $L=0$ and $R=5^m-1$, or in 5-ary:
$$
L = 0
\qquad\text{and}\qquad
R=
\underbrace{4\,\,4\,\,4\,\,\cdots\,\,4}_{m\text{ times}}.
$$
Hence $L$ and $R$ correctly represent the starting tape configuration
of $A$ on input $0^m$:
\begin{center}
\footnotesize
$
\begin{array}{ccccccccclc}
\cdots&l_1 &l_0 &r_0&r_1&r_2& \cdots &r_{m-1}&r_m &r_{m+1}&\cdots \\ \hline
\cdots&\bla&\bla& 0 & 0 & 0 & \cdots & 0     &\bla&\bla&\cdots \\ \hline
      &    &    & \uparrow & 
\end{array}
$
\end{center}
(since symbol $0$ maps to~$4$) and $B$ is ready to start a faithful
simulation of $A$'s steps. The details of this simulation are
straightforward and are given in the Appendix.

It is important to note the range of the values assumed by the
counters. The second part of the initialization phase can be
implemented so that it never assigns to a counter a value greater than
the starting value of $J$, which is $\lg_{30} n$. Then, in the
simulation phase, the behaviour of $A$ (the tape starts with $m$ $0$'s
and always contains exactly $m$ non-blank symbols) and the selection
of the symbol codes ($0$~gets the largest code) are such that the
initial value $5^m-1$ of $R$ upper bounds all possible values that may
appear in $B$'s counters.  Hence, since $5^m-1<\lg_{30} n$, \emph{the
entire computation of $B$ after the first part of its initialization
phase keeps all values of all counters at or below} $\lg_{30} n$. In
particular, \emph{after the first part of the initialization phase,
all attempts to increment a counter are successful}. This will prove
crucial in the next section.

\paragraph{The third machine.}
$C$ is a $\dca_2$ that simulates the behaviour of~$B$. If $U$, $V$ are
its two counters, then $V$ is auxiliary for performing operations on
$U$, while $U$ encodes the contents of the counters of $B$: whenever
$L$, $R$, $J$ contain $l$, $r$, $j$ respectively, $U$ contains
$2^l3^r5^j$.

The automaton starts by computing into $U$ the product
$30^t=2^t3^t5^t$, where $t=\lg_{30}n$ (if $n=0$, $C$ rejects, exactly
as $B$ would do). It then removes all $2$'s and $3$'s from this
product, so that $U$ becomes $2^03^05^{\lg_{30}n}$, which correctly
encodes the values of the counters of $B$ \emph{right after the first
part of its initialization phase}. At this point $C$ starts a faithful
simulation of the steps of~$B$.

The simulation is straightforward: every attempt of $B$ to increase
[decrease] a counter is simulated by an attempt of $C$ to multiply
[divide] $U$ by the corresponding prime ---see the Appendix for the
details. The only subtlety involves failure during increment attempts:
How does the simulation make sure that \emph{an attempt of $B$ to
increase a counter fails} iff \emph{the corresponding attempt of $C$
to multiply $U$ fails}?\footnote{Note that this condition is necessary
for a successful simulation but is not met in some obvious way. In $B$
all three counters have the same upper bound ($B$'s input), whereas in
$C$ their representations have different upper bounds: the base~2,
base~3, and base~5 logarithms of the upper bound for $U$ ($C$'s
input).} The crucial observation (from the previous section) is that,
since we are after the first part of $B$'s initialization phase,
\emph{every attempt of $B$ to increment a counter is successful}
(which makes the left-to-right direction of the equivalence above
vacuously true) and \emph{produces a value at or below~$t$} (which
implies the value of $U$ never exceeds its original value $2^t3^t5^t$,
making the right-to-left direction of the equivalence above vacuously
true, as well). Hence, during the simulation, all multiplication
attempts of $C$ are successful and simulate only successful increment
attempts of~$B$.

\paragraph{The final machine.}
$M'$ is a $\utdfa_2$ that, on input $n$, simulates $C$ on input $n+1$
mimicking the counter operations by head moves
(cf.~Lemma~\ref{lem:dca-vs-utdfa}). 

\bigskip\noindent
This concludes the (informal) definitions of all
four machines of our reduction. It should be clear that $A$, $B$, and
$C$, are all good, so that $M'$ is good, as well. Moreover, a
description $z'$ of $M'$ can clearly be computed out of the
description $z$ of~$M$.

\section{The second reduction}
\label{sec:from-eot-to-ess_k}

Our goal in this section is to find an algorithm that, on input a
description $z$ of a terminating $\utdfa_2$ $M$ whose language is
guaranteed to be either empty or threshold, produces a description
$z'$ of a $\tdfa_{k+1}$ $M'$ such that
\begin{equation} \label{eq:reduction-condition-2}
\begin{split}
(M)=\emptyset 
&\implies 
M'\text{ has no equivalent }\tdfa_k,
\\
(M)\text{ is a threshold language} 
&\implies 
M'\text{ has equivalent $\tdfa_k$s}.
\end{split}
\end{equation}
Following the idea of~\cite[Theorem~7]{ku03}, we will need an
arbitrary $\tdfa_{k+1}$ that has no equivalent $\tdfa_k$. So, fix
$D$ to be such a $\tdfa_{k+1}$ and $z_D$ to be a description of it;
also let $\ggS$ be $D$'s input alphabet.

Our algorithm is simple. It uses $z$ and $z_D$ to construct a
description $z'$ for the $\tdfa_{k+1}$ $M'$ that, on input
$x\in\ggS\s$, follows the instructions:

\smallskip
\qquad
1.~Simulate $M$ on $x$;  if $M$ accepts, accept; otherwise, go to
Step~2. 

\qquad
2.~Simulate $D$ on $x$  (and accept, reject, or loop accordingly).
\smallskip

\noindent
Note that in the first simulation $M'$ treats all symbols of $\ggS$
the same. Also, this simulation \emph{is} possible, since $k+1\geq 2$.
Moreover, when $M$ rejects, $M'$ \emph{can} move to Step~2, since
$M$ never rejects by looping. Overall, $M'$ is well-defined.

In addition, the language of $M'$ is clearly
$\lang{M}\cup\lang{D}$. Hence, when $\lang{M}=\emptyset$, then
$\lang{M'}$ is just $\lang{D}$; when $(M)$ is a threshold language
with threshold $l$, then
$$
(M')=
\{ x\in\lang{D} \mid \text{length}(x) < l   \}
\;\;\cup\;\;
\{ x\in\ggS\s   \mid \text{length}(x)\geq l \},
$$
which is cofinite, and hence regular. In total, 
\begin{equation} \label{eq:M-vs-M'}
\lang{M}=\emptyset 
\Rightarrow
\lang{M'}=\lang{D}
\quad\amp\quad
\lang{M}\text{ is threshold} 
\Rightarrow
\lang{M'}\text{ is regular},
\end{equation}
which implies~(\ref{eq:reduction-condition-2}), by the selection of
$D$ and the inclusion of regular languages in
$\lang{\tdfa_1}\subseteq\lang{\tdfa_k}$.  Hence, our algorithm can
safely output~$z'$.

\paragraph{The unary and nondeterministic cases.}
By appropriately changing the selection of $z_D$, the same algorithm
establishes that $\eot\leq\ess_k$, even when $\ess_k$ stands for the
inadequacy of $k$ heads for $\utdfa_{k+1}s$ [$\tnfa_{k+1}s$,
$\utnfa_{k+1}s$].  More specifically, we now select $z_D$ to be a
description of a $\utdfa_{k+1}$ [$\tnfa_{k+1}$, $\utnfa_{k+1}$] $D$
that has no equivalent $\utdfa_k$ [$\tnfa_k$, $\utnfa_k$]. Then, $M'$
is a well-defined automaton of the same type as $D$ (note that $M'$
has enough resources to simulate each of $M$, $D$) that
satisfies~(\ref{eq:M-vs-M'}), and hence the version
of~(\ref{eq:reduction-condition-2}) for $\utdfa_k$s [$\tnfa_k$s,
$\utnfa_k$s].

\section{Conclusion}
\label{sec:concl}

We have proved that one extra head allows non-recursively greater
succinctness in describing languages of $\l$ by multihead automata, be
they deterministic or nondeterministic, unary or not. It is fairly
easy to draw similar conclusions for other types of (pebble or
counter) automata of comparable power.

\newpage
\appendix

\section*{Appendix}

\paragraph{Computing with a multicounter automaton.}
In order to facilitate the description of the behaviour of $\mdca$s,
we introduce some `program' notation.  First, the two atomic
operations, the attempt to decrease a counter $X$ and the attempt to
increase it, are denoted respectively by
$$
X \stackrel{f}{\longleftarrow} X-1
\qquad\qquad\text{and}\qquad\qquad
X \stackrel{f}{\longleftarrow} X+1,
$$
where, in each case, flag $f$ is set to true iff the attempt
succeeds. Then, the compound operation of setting $X$ to~$0$, denoted
by $X\longleftarrow 0$, can be described by the line
\begin{equation} \label{eq:set-X-to-0}
\texttt{repeat }\;
X \stackrel{f}{\longleftarrow} X-1\;
\texttt{ until $\neg f$}.
\end{equation}
If a second counter $Y$ is present, we can tranfer the contents of $Y$
into $X$: we set $X$ to~$0$, then repeatedly decrease $Y$ and
increase $X$ until $Y$ is~0. We denote this by
$$
(X,Y)\longleftarrow(Y,0),
$$
and describe it by a line similar to~(\ref{eq:set-X-to-0}).  Changing
how fast $X$ increases as $Y$ decreases, we can multiply/divide $Y$
into $X$ by any constant $a$. We denote these operations by
$$
(X,Y)
\stackrel{f}{\longleftarrow}
(aY,0)
\qquad\text{and}\qquad
(X,Y)
\stackrel{r}{\longleftarrow}
\bigl(\bigl\lfloor\tfrac{Y}{a}\bigr\rfloor,0\bigr).
$$
For the first one, note that $aY>n$ implies one of the attempts to
increase $X$ will fail; in that case, we restore the original value of
$Y$ returning $X$ to~0, and set flag $f$ to false. In the second
operation, we also find the remainder and return it in~$r$.

At a higher level, we can attempt to multiply $Y$ by a constant $a$
(into $Y$) using $X$ as an auxiliary counter and making sure $Y$
changes only if the operation succeeds:
$$
(X,Y)
\stackrel{f}{\longleftarrow}
(aY,0);
\quad
\texttt{if $f$ then }
(Y,X)
\longleftarrow
(X,0).
$$
Division (with remainder) can be done similarly.  We denote the two
operations~by
\begin{equation} \label{eq:mult-div}
Y 
\stackrel{\;\,f,X}{\longleftarrow}
aY
\qquad\qquad\text{and}\qquad\qquad
Y 
\stackrel{\;\,r,X}{\longleftarrow}
\bigl\lfloor\tfrac{Y}{a}\bigr\rfloor.
\end{equation}
Now, we can set $Y$ to the largest power of $a$ that can fit in $n$:
\begin{equation} \label{eq:implement-log}
Y\longleftarrow 0;
\;
Y\stackrel{f}{\longleftarrow} Y+1;
\;
\texttt{if $f$ then repeat}\;\;
Y \stackrel{\;\,g,X}{\longleftarrow} aY\;
\texttt{until $\neg g$}
\end{equation}
(note that this fails iff $n=0$).  We denote~(\ref{eq:implement-log})
by (remember that $\lg_a n = \lfloor\log_a n\rfloor$):
$$
Y \stackrel{\;\,f,X}{\longleftarrow} a^{\lg_a n}.
$$
If a third counter $Z$ is present, we can
modify~(\ref{eq:implement-log}) to also count (in~$Z$) the number of
iterations performed. This gives us a way to try to calculate $\lg_a
n$ (failing iff $n=0$):
$$
Z 
\stackrel{\;f,X,Y}{\longleftarrow} 
\lg_a n.
$$
In another variation, we can modify the multiplication
in~(\ref{eq:mult-div}) so that the success of the operation depends on
the contents of $Z$ (as opposed to~$n$). We write this as
$$
Y 
\stackrel{\;\,f,X,Z}{\longleftarrow}
aY,
$$
meaning that, using $X$ as auxiliary and without affecting $Z$: if
$aY\leq Z$, then $Y$ is set to $aY$; otherwise, $Y$ is
unaffected.\footnote{To implement this, we repeatedly decrease $Y$,
increase $X$, and decrease $Z$ by $a$. \emph{If $Z$ becomes~0 before
$Y$}, then $aY>Z$ and the operation should fail: we restore the
original values of $Y$ and $Z$ by repeatedly decreasing $X$,
increasing $Y$, and increasing $Z$ by $a$, until $X$
becomes~0. \emph{Otherwise}, $aY\leq Z$ and the operation will
succeed: we copy the correct value to $Y$ and restore the value of $Z$
by repeatedly decreasing $X$ and increasing each of $Y$, $Z$ by~$a$,
until $X$ becomes~0. Note that \emph{none of the counters ever assumes
a value greater than the original value
of~$Z$}.\label{note:implement-bounded-mult}} Then, the following
variant of~(\ref{eq:implement-log})
$$
Z\stackrel{f}{\longleftarrow} Z-1;
\;\;
\texttt{if $f$ then }
\{
Z\stackrel{\texttt{t}}{\longleftarrow} Z+1;
\;\;
\texttt{repeat }
Y \stackrel{\;\,g,X,Z}{\longleftarrow} aY
\;\;
\texttt{until $\neg g$}
\}
$$
implements the attempt to set $Y$ to the largest power of $a$ that is
at most $Z$, using $X$ as auxiliary and leaving $Z$ unaffected
(failing iff $Z$ is~0).\footnote{Note the use of \texttt{t} in the
place of a flag, indicating that the action is guaranteed to succeed.}
We denote this one by
$$
Y \stackrel{\;\,f,X}{\longleftarrow} a^{\lg_a Z}.
$$
It is important to note that \emph{during this operation no counter
ever assumes a value greater than the original value of $Z$}
(cf. Footnote~\ref{note:implement-bounded-mult}).

\paragraph{The computation of $B$.}
In the program notation from the previous section, the two parts of
the initialization phase of $B$ are given respectively by the two
lines
\begin{gather*} 
L
\stackrel{\;\,f,R,J}{\longleftarrow}
\lg_{30} n;
\;\;
\texttt{if $\neg f$ then reject else }
(J,L)\longleftarrow (L,0),
\\
R \stackrel{\;\,f,L}{\longleftarrow} 5^{\lg_5 J};
\;
\texttt{if $\neg f$ then reject else }
\{
R \stackrel{\texttt{t}}{\longleftarrow} R-1;
\;
L \longleftarrow 0
\}.
\end{gather*}
For the simulation of $A$, the automaton remembers the current state
of $A$ in its finite memory. To find the code $r_0$ of the symbol read
by the head of $A$, it computes:
$$
R 
\stackrel{\;\,r_0,J}{\longleftarrow}
\bigl\lfloor\tfrac{R}{5}\bigr\rfloor.
$$
Then, if $s$ is the code of the symbol to be written on the tape, it
computes
$$
L
\stackrel{\;\,\texttt{t},J}{\longleftarrow}
5L;
\;\;
\texttt{repeat $s$ times:}\;\;
L\stackrel{\texttt{t}}{\longleftarrow} L+1
$$
to simulate writing this symbol and moving to the right; it computes
\begin{multline*}
\qquad\qquad
R
\stackrel{\;\,\texttt{t},J}{\longleftarrow}
5R;
\;\;
\texttt{repeat $s$ times:}\;\;
R\stackrel{\texttt{t}}{\longleftarrow} R+1;
\\
L 
\stackrel{\;\,l_0,J}{\longleftarrow}
\bigl\lfloor\tfrac{L}{5}\bigr\rfloor;
\;\;
R
\stackrel{\;\,\texttt{t},J}{\longleftarrow}
5R;
\;\;
\texttt{repeat $l_0$ times:}\;\;
R\stackrel{\texttt{t}}{\longleftarrow} R+1
\qquad\qquad
\end{multline*}
to simulate writing this symbol and moving to the left.\footnote{Some
extra care is necessary when $L=0$ or $R=0$, in which case the read
head is at the leftmost or rightmost input symbol. We omit the
details.} Note the $\texttt{t}$ on the increment attempts, a reminder
that the attempts are guaranteed to be successful.

\paragraph{The computation of $C$.}
The automaton starts with the computation 
$$
U \stackrel{\;\,f,V}{\longleftarrow} 30^{\lg_{30} n};
\;\;
\texttt{if $\neg f$ then reject}.
$$
It then removes from $U$ all $2$'s by repeated divisions
($ 
U
\stackrel{\;\,r,V}{\longleftarrow}
\bigl\lfloor\tfrac{U}{2}\bigr\rfloor 
$), 
until a non-zero remainder is returned, which implies there were no
$2$'s in $U$ before the last division; then a correction 
($
U 
\stackrel{\;\,\texttt{t},V}{\longleftarrow}
2U;
\;\;
U \stackrel{\texttt{t}}{\longleftarrow} U+1
$)
undoes the damage caused by this last division. Subsequently, a
similar computation removes all $3$'s.

For the simulation part, when $B$ attempts to decrease~$J$
($
J \stackrel{f}{\longleftarrow} J-1
$),
$C$ divides $U$ by~5; if there is no remainder, the division simulated
a successful decrement; otherwise, the simulated attempt failed
and $C$ restores $U$'s initial value:
\begin{multline*}
U 
\stackrel{\;\,r,V}{\longleftarrow}
\bigl\lfloor\tfrac{U}{5}\bigr\rfloor;
\;\;
\texttt{if $r=0$ then $f\longleftarrow \texttt{true}$ else}
\\
\{
f\longleftarrow \texttt{false};
\;\;
U 
\stackrel{\;\,\texttt{t},V}{\longleftarrow}
5U;
\;\;
\texttt{repeat $r$ times:}\;\;
U \stackrel{\texttt{t}}{\longleftarrow} U+1
\}.
\end{multline*}
When $B$ attempts to increase $J$ 
($
J \stackrel{\texttt{t}}{\longleftarrow} J+1
$),
$C$ just multiplies $U$ by $5$ 
($
U \stackrel{\;\,\texttt{t},V}{\longleftarrow} 5U
$).%
\footnote{As explained in Section~\ref{sec:from-looping-to-eot}, both
the increment and the multiplication are guaranteed to be successful,
hence the $\texttt{t}$ reminder.} The attempts to increase or decrease
$L$ or $R$ are handled similarly, with $2$ or $3$ instead of~$5$.

\bibliographystyle{plain}
\bibliography{string,automata}

\end{document}